\documentclass[10pt,conference]{IEEEtran}
\usepackage{epsfig,graphicx,subfigure,psfrag,amsmath,cases,bm}
\usepackage{latexsym,amssymb,amsmath,epsfig,subfigure,algorithm,mathtools}
\usepackage{algorithmic}
\usepackage{color}
\usepackage{url}
\usepackage{scrtime}
\usepackage{stfloats}
\author{\IEEEauthorblockN {Dongfang Xu\IEEEauthorrefmark {3}, Xianghao Yu\IEEEauthorrefmark {3}, Vahid Jamali\IEEEauthorrefmark {3}, Derrick Wing Kwan Ng\IEEEauthorrefmark {4}, and Robert Schober\IEEEauthorrefmark {3}}

\IEEEauthorrefmark {3}Friedrich-Alexander-University
Erlangen-N\"urnberg, Germany,
\IEEEauthorrefmark {4}The University
of New South Wales, Australia

}

\newtheorem{T-Prob}{Transformed Problem}

\newcounter{TempEqCnt}

\DeclareMathOperator{\mino}{minimize}

\newcommand{\qed}{\hfill \ensuremath{\blacksquare}}
\newtheorem{Remark}{Remark}

\allowdisplaybreaks

\title{Resource Allocation for Large IRS-Assisted SWIPT Systems with Non-linear Energy Harvesting Model}

\begin{document}
\maketitle
\begin{abstract}
In this paper, we investigate resource allocation algorithm design for large intelligent reflecting surface (IRS)-assisted simultaneous wireless information and power transfer (SWIPT) systems. To this end, we adopt a physics-based IRS model that, unlike the conventional IRS model, takes into account the impact of the incident and reflection angles of the impinging electromagnetic wave on the reflected signal. To facilitate efficient resource allocation design for large IRSs, we employ a scalable optimization framework, where the IRS is partitioned into several tiles and the phase shift elements of each tile are jointly designed to realize different transmission modes. Then, the beamforming vectors at the base station (BS) and the transmission mode selection of the tiles of the IRS are jointly optimized for minimization of the BS transmit power taking into account the quality-of-service requirements of both non-linear energy harvesting receivers and information decoding receivers. For handling the resulting non-convex optimization problem, we apply a penalty-based method, successive convex approximation, and semidefinite relaxation to develop a computationally efficient algorithm which asymptotically converges to a locally optimal solution of the considered problem. Our simulation results show that the proposed scheme enables considerable power savings compared to two baseline schemes. Moreover, our results also illustrate that the advocated physics-based model and scalable optimization framework for large IRSs allows us to strike a balance between system performance and computational complexity, which is vital for realizing large IRS-assisted communication systems.
\end{abstract}
\section{Introduction}
Recently, radio frequency (RF) transmission enabled simultaneous wireless information and power transfer (SWIPT) has emerged as a promising technique for facilitating sustainable operation of energy-constrained communication systems \cite{wong2017key}, \cite{6781609}. However, since the path loss is proportional to the transmission distance, a significant amount of energy can be harvested only within a small area around the transmitter. Moreover, due to the randomness of wireless channels, the performance of SWIPT systems can be significantly degraded when the radio propagation environment is unfavorable. One promising solution to overcome these obstacles is the application of intelligent reflecting surfaces (IRSs). In particular, an IRS consists of numerous energy-efficient phase shift elements where each element can reflect a received electromagnetic wave and apply a desired phase shift \cite{cui2014coding}. By intelligently adapting the IRS elements to the channel conditions, the reflected signals can be constructively combined at desired locations creating a favorable radio propagation environment for performance improvement \cite{8910627}. As a result, several works have proposed the application of  IRSs in SWIPT systems to enhance system performance. In particular, the authors of \cite{9148892} studied the joint design of the beamforming vector at the base station (BS) and the phase shift pattern of the IRS for maximization of the minimum harvested energy among all the energy harvesting receivers (ERs) in an IRS-aided SWIPT system. In \cite{pan2020intelligent}, the authors considered an IRS-enabled multiple-input multiple-output (MIMO) SWIPT system and developed an alternating optimization (AO)-based algorithm for maximization of the system spectral efficiency while providing reliable wireless power transfer service to multiple ERs. However, the authors of \cite{9148892} and \cite{pan2020intelligent} assumed that the IRS can reflect the signals originating from different directions with unit gain and adopted a linear energy harvesting (EH) model.
\par
In practice, depending on the incident angle, the reflection angle, and the polarization of the electromagnetic wave, the IRS reflects the signals incident from different directions with different gains \cite{najafi2020physics}. Moreover, in free space propagation environments, the path loss of the BS-IRS-receiver link is much larger than that of the direct link for far-field scenarios \cite{di2020analytical}. It has been shown in \cite{najafi2020physics} that for the path loss of the IRS-assisted link to be the same as that of the unobstructed direct link, the IRS should be equipped with hundreds of phase shift elements for typical system parameters. Moreover, the resource allocation algorithms proposed in \cite{9148892} and \cite{pan2020intelligent} optimize the phase shift of each individual IRS element which entails a high complexity. Hence, these algorithms may not be efficient and suitable for online resource allocation design for large IRS-assisted SWIPT systems in practice. Furthermore, the authors of \cite{9148892} and \cite{pan2020intelligent} adopted an overly-simplified EH model, which assumes that the harvested power of the ERs increases linearly with the received RF power. However, according to measurements \cite{valenta2014harvesting}, \cite{le2008efficient}, the linear EH model is only accurate when the received RF power does not vary significantly. In fact, due to the combination of the signals from the direct path and the reflected path, the received RF power at the ERs in IRS-assisted SWIPT systems usually experiences a larger dynamic range than in conventional SWIPT systems. Therefore, the resource allocation algorithm designs proposed in \cite{9148892} and \cite{pan2020intelligent} may not be able to effectively enhance the performance of practical large IRS-assisted SWIPT systems.
\par
Motivated by the above discussion, in this paper, we investigate resource allocation algorithm design for large IRS-assisted SWIPT systems. We adopt the physics-based IRS model which has been recently developed in \cite{najafi2020physics} and characterize the IRS in terms of a finite number of phase shift configurations, which we refer to as transmission modes. Moreover, we adopt the non-linear EH model for ERs from \cite{boshkovska2015practical}. Based on these IRS and EH models, we minimize the total transmit power of the IRS-assisted SWIPT system by jointly designing the BS beamforming and IRS transmission mode selection policy. Due to the binary constraint introduced by the transmission mode selection and the coupled optimization variables, determining the globally optimal solution of the resulting optimization problem is generally intractable. Therefore, in this paper, by capitalizing on a penalty-based method, successive convex approximation (SCA), and semidefinite relaxation, we develop a computationally efficient algorithm which asymptotically converges to a locally optimal solution of the considered problem.
\par
\textit{Notation:} In this paper, boldface lower case and boldface capital letters denote vectors and matrices, respectively. $\mathbb{C}^{N\times M}$ denotes the space of $N\times M$ complex-valued matrices. $\mathbb{H}^{N}$ denotes the set of all $N$-dimensional complex Hermitian matrices. $\mathbf{I}_{N}$ refers to the $N\times N$ identity matrix. $|\cdot|$ and $||\cdot||_2$ denote the absolute value of a complex scalar and the $l_2$-norm of a vector, respectively. $\mathbf{A}^H$ stands for the conjugate transpose of matrix $\mathbf{A}$. $\mathbf{A}\succeq\mathbf{0}$ indicates that $\mathbf{A}$ is a positive semidefinite matrix. $\mathrm{Rank}(\mathbf{A})$ and $\mathrm{Tr}(\mathbf{A})$ denote the rank and the trace of matrix $\mathbf{A}$, respectively. $\mathcal{E}\left \{ \cdot \right \}$ denotes statistical expectation. $\sim$ and $\overset{\Delta }{=}$ stand for ``distributed as'' and ``defined as'', respectively. The distribution of a circularly symmetric complex Gaussian random variable with mean $\mu$ and variance $\sigma^2$ is denoted by $\mathcal{CN}(\mu ,\sigma^2)$. $\mathbf{x}^*$ denotes the optimal value of optimization variable $\mathbf{x}$. 
\section{System Model}
\begin{figure}[t]
\vspace*{0mm}
\centering
\includegraphics[width=3.4in]{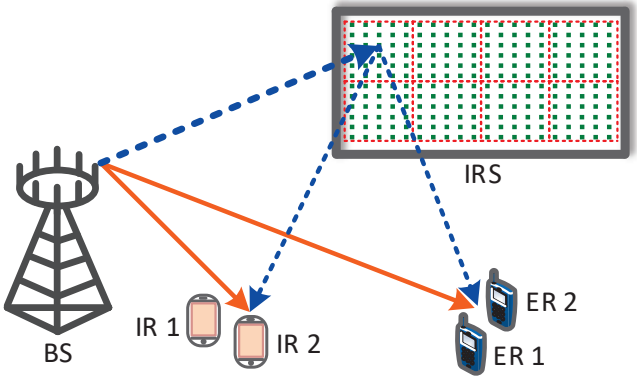} \vspace*{0mm}
\caption{An IRS-aided SWIPT system comprising one multi-antenna base station, $K=2$ information decoding receivers, and $J=2$ energy harvesting receivers. To facilitate efficient online design, the large IRS is partitioned into $T=8$ tiles of equal size, as indicated by the red-colored dotted boxes.}
\label{system_model}\vspace*{0mm}
\end{figure}
The considered IRS-assisted SWIPT system comprises a BS, $K$ information decoding receivers (IRs), and $J$ ERs, cf. Figure \ref{system_model}. In particular, the BS is equipped with $N_{\mathrm{T}}$ antennas while all receivers are single-antenna devices. To enhance the performance of the considered system, a large IRS is deployed
to assist in the SWIPT from the BS to the two sets of receivers. 
Specifically, we consider a large rectangular IRS comprising $M$ phase shift elements. Moreover, we adopt the scalable physics-based IRS model recently developed in \cite{najafi2020physics} to facilitate computationally efficient resource allocation algorithm design for large IRSs. In particular, we partition the large IRS into $T$ tiles of equal size and then optimize each tile in two stages, namely an offline design stage and an online optimization stage. In the offline design stage, we generate a codebook by jointly designing the phase shift elements of each tile for the support of $S$ different transmission modes\footnote{Depending on the requirements, the $S$ transmission modes can be designed to realize reflection along different directions or/and to ensure the coherent combination of the reflected signals originating from different tiles, see \cite{najafi2020physics} for more details.}, where each transmission mode corresponds to a certain configuration of the phase shifts that the tile applies to the incident signals. In the online optimization stage, from the transmission mode set, we choose the best transmission mode for each tile such that the performance is maximized. In the following, we assume the transmission mode set of each tile has already been generated in the offline design stage, using e.g. the approach in \cite{najafi2020physics}, and focus on the online optimization stage. Besides, we define sets $\mathcal{J}=\left \{1,\cdots ,J \right \}$ and $\mathcal{K}=\left \{1,\cdots ,K \right \}$ to collect the indices of the ERs and IRs, respectively.
\par
In a given scheduling time slot, the BS transmit signal is given by
\begin{equation}
    \mathbf{x}=\underset{k\in\mathcal{K} }{\sum }\mathbf{w}_kd_k^{\mathrm{I}}+\underset{j\in\mathcal{J} }{\sum }\mathbf{v}_jd_j^{\mathrm{E}},
\end{equation}
where $\mathbf{w}_k\in \mathbb{C}^{\mathit{N}_{\mathrm{T}}\times 1}$ and $d_k^{\mathrm{I}}\in \mathbb{C}$ denote the beamforming vector for IR $k$ and the corresponding information symbol while $\mathbf{v}_j\in \mathbb{C}^{\mathit{N}_{\mathrm{T}}\times 1}$ and $d_j^{\mathrm{E}}\in \mathbb{C}$ denote the beamformer for ER $j$ and the corresponding energy signal. Without loss of generality, we assume $\mathcal{E}\{\left |d_k^{\mathrm{I}} \right|^2\}=1$, $\forall\mathit{k} \in \mathcal{K}$, and $\mathcal{E}\{\left |d_j^{\mathrm{E}} \right|^2\}=1$, $\forall\mathit{j} \in \mathcal{J}$.
\par
The received signals at IR $k$ and ER $j$ are given by
\begin{eqnarray}
    y_k^{\mathrm{I}}&\hspace*{-3mm}=\hspace*{-3mm}&\underset{\substack{s\in\mathcal{S},\\t\in\mathcal{T}\cup \left \{ 0 \right \}}}{\sum }b_{s,t}\mathbf{h}^H_{s,t,k}\left ( \underset{r\in\mathcal{K} }{\sum }\mathbf{w}_rd_r^{\mathrm{I}}+\underset{j\in\mathcal{J} }{\sum }\mathbf{v}_jd_j^{\mathrm{E}} \right )+n_k^{\mathrm{I}},\\
    y_j^{\mathrm{E}}&\hspace*{-3mm}=\hspace*{-3mm}&\underset{\substack{s\in\mathcal{S},\\t\in\mathcal{T}\cup \left \{ 0 \right \}}}{\sum }b_{s,t}\mathbf{g}^H_{s,t,j}\left ( \underset{k\in\mathcal{K} }{\sum }\mathbf{w}_kd_k^{\mathrm{I}}+\underset{i\in\mathcal{J} }{\sum }\mathbf{v}_id_i^{\mathrm{E}} \right )+n_j^{\mathrm{E}},
\end{eqnarray}
respectively, where sets $\mathcal{S}=\left \{ 1,\cdots ,S \right \}$ and $\mathcal{T}=\left \{ 1,\cdots ,T \right \}$ collect the indices of the transmission modes and tiles, respectively. Furthermore, $b_{s,t}$ is a binary optimization variable with $b_{s,t}=1$ if tile $t$ employs transmission mode $s$, $\forall s\in\mathcal{S}$, $\forall t\in\mathcal{T}$; otherwise it is equal to zero. Variables $n_k^{\mathrm{I}}\sim\mathcal{CN}(0,\sigma_{\mathrm{I}_k}^2)$ and $n_j^{\mathrm{E}}\sim\mathcal{CN}(0,\sigma_{\mathrm{E}_j}^2)$ denote the additive white Gaussian noises at IR $k$ and ER $j$, respectively. For tile $t$ employing transmission mode $s$, the effective end-to-end channel between the BS and IR $k$, i.e., $\mathbf{h}_{s,t,k}$, is given by \cite{najafi2020physics}
\begin{equation}
\label{channel}
    \mathbf{h}_{s,t,k}^H=\mathbf{a}_{\mathrm{R}_k}^H\mathbf{C}_{\mathrm{R}_k}\mathbf{R}_{s,t,k}\mathbf{C}_{\mathrm{T}}\mathbf{D}_{\mathrm{T}},~~\forall t\in\mathcal{T}.
\end{equation}
Here, $\mathbf{a}_{\mathrm{R}_k}\in\mathbb{C}^{L_{\mathrm{R}_k}\times 1}$ denotes the receive steering vector at IR $k$ evaluated at the angles-of-arrival (AoAs) of the paths between the IRS and IR $k$. $\mathbf{D}_{\mathrm{T}}\in\mathbb{C}^{L_{\mathrm{T}}\times N_{\mathrm{T}}}$ is a matrix whose rows are the transmit steering vectors at the BS evaluated at the angles-of-departure (AoDs) of the paths between the IRS and the BS. $L_{\mathrm{T}}$ and $L_{\mathrm{R}_k}$ denote the numbers of scatterers of the BS-IRS and IRS-IR $k$ links, respectively. Moreover, $\mathbf{C}_{\mathrm{T}}\in\mathbb{C}^{L_{\mathrm{T}}\times L_{\mathrm{T}}}$ and $\mathbf{C}_{\mathrm{R}_k}\in\mathbb{C}^{L_{\mathrm{R}_k}\times L_{\mathrm{R}_k}}$ are diagonal matrices containing the channel coefficients of the BS-IRS and IRS-IR $k$ paths\footnote{We note that $\mathbf{C}_{\mathrm{T}}$ and $\mathbf{C}_{\mathrm{R}_k}$ capture the joint impact of path loss, shadowing, and small-scale fading on the corresponding paths.}, respectively. Furthermore, matrix $\mathbf{R}_{s,t,k}\in\mathbb{C}^{L_{\mathrm{R}_k}\times L_{\mathrm{T}}}$ denotes the response function of tile $t$ applying the $s$-th transmission mode evaluated at the AoAs and AoDs of the BS-IRS-IR $k$ link. For given incident angles, reflection angles, and polarization of the electromagnetic wave, $\mathbf{R}_{s,t,k}$ can be calculated according to \cite[Propositions 1 and 2]{najafi2020physics}. On the other hand, for convenience of presentation, we define the channel vector of the direct link between the BS and IR $k$ via a virtual tile $t=0$ as
\begin{equation}
    \mathbf{h}_{s,0,k}^H=\mathbf{a}_{\mathrm{D}_k}^H\mathbf{C}_{\mathrm{D}_k}\mathbf{D}_{\mathrm{D}},
\end{equation}
where $\mathbf{a}_{\mathrm{D}_k}\in\mathbb{C}^{L_{\mathrm{D}_k}\times 1}$, $\mathbf{D}_{\mathrm{D}}\in\mathbb{C}^{L_{\mathrm{D}_k}\times N_{\mathrm{T}}}$, and $\mathbf{C}_{\mathrm{D}_k}\in\mathbb{C}^{L_{\mathrm{D}_k}\times L_{\mathrm{D}_k}}$ are the respective components of the direct path between the BS and IR $k$. Here, $L_{\mathrm{D}_k}$ denotes the number of scatterers of the BS-IR $k$ link. We note that the effective end-to-end channels of the reflected BS-IRS-ER $j$ link and the direct BS-ER $j$ link, i.e., $\mathbf{g}_{s,t,j}$ and $\mathbf{g}_{s,0,j}$, are defined in a similar manner as $\mathbf{h}_{s,t,k}$ and $\mathbf{h}_{s,0,k}$, respectively.
Furthermore, to determine the maximum achievable performance, similar to \cite{8941080}, \cite{yu2020power}, we assume that the perfect channel state information (CSI) of the system is available at the BS. For notational simplicity, we define a new index set $\widehat{\mathcal{T}}=\mathcal{T}\cup \left \{ 0 \right \}$.
\par
\begin{Remark}
In this paper, we adopt the physics-based channel model in \cite{najafi2020physics} for IRS-assisted SWIPT systems which facilitates scalable optimization for large IRSs. Assuming the effective end-to-end channels in \eqref{channel} are known, we aim to select the best transmission mode for each tile from the given transmission mode set. As a result, the computational complexity of the proposed resource allocation algorithm design does not scale with the number of the IRS elements $M$, but with the numbers of tiles and transmission modes, i.e., $T$ and $S$. Hence, by properly adjusting the number of tiles and the number of transmission modes, the computational complexity of the online design of large IRSs becomes affordable for practical IRS-assisted SWIPT systems.
\end{Remark}
\par
Moreover, several existing works adopted a linear EH model for IRS-assisted SWIPT systems \cite{9148892}, \cite{pan2020intelligent}. However, this model is not accurate as the RF energy conversion efficiency depends on the input RF power level of the EH circuit. To capture this effect, in this paper, we adopt the non-linear EH model proposed in \cite{boshkovska2015practical}. In particular, the energy harvested by ER $j$, i.e., $\Upsilon_j^{\mathrm{EH}}$, is given by 
\begin{eqnarray}
    \Upsilon_j^{\mathrm{EH}}&\hspace*{-2mm}=\hspace*{-2mm}&\frac{\Lambda_j-a_j\Xi_j}{1-\Xi_j}, ~~\Xi_j=\frac{1}{1+\mathrm{exp}(\varrho _jc_j)},\\
    \Lambda_j&\hspace*{-2mm}=\hspace*{-2mm}&\frac{a_j}{1+\mathrm{exp}\left(-\varrho _j\left ( P^{\mathrm{ER}}_j-c_j \right )\right)},
\end{eqnarray}
where $\Lambda_j$ is the logistic function for a given received RF power $P^{\mathrm{ER}}_j$ at ER $j$. Moreover, constant $\Xi_j$ is imposed to ensure a zero-input/zero-output response. Besides, $a_j$, $c_j$, and $\varrho _j$ are
constant parameters determined by the employed EH circuit \cite{boshkovska2015practical}.
\section{Optimization Problem Formulation}
In this section, the resource allocation optimization problem is formulated for the considered system. However, first the adopted performance metrics are defined.
\par
\begin{figure*}[t]
\setcounter{TempEqCnt}{\value{equation}} 		 
	\setcounter{equation}{\value{equation}}
\begin{eqnarray}
\label{SINR}
    \hspace*{-6mm}\Gamma_k&\hspace*{-3mm}=\hspace*{-3mm}&\frac{\underset{\substack{s,p\in\mathcal{S},\\t,q\in\widehat{\mathcal{T}}}}{\sum }b_{s,t}b_{p,q}\mathbf{h}^H_{s,t,k} \mathbf{w}_k\mathbf{w}^H_k\mathbf{h}_{p,q,k}}{\hspace*{-1mm}\underset{\substack{s,p\in\mathcal{S},\\t,q\in\widehat{\mathcal{T}}}}{\sum }\hspace*{-1mm}b_{s,t}b_{p,q}\mathbf{h}^H_{s,t,k}\left (  \underset{r\in\mathcal{K}\setminus \left \{ k \right \} }{\sum }\hspace*{-1mm}\mathbf{w}_r\mathbf{w}^H_r\hspace*{-1mm}+\hspace*{-1mm}\underset{j\in\mathcal{J}}{\sum }\mathbf{v}_j\mathbf{v}^H_j \hspace*{-1mm}\right )\mathbf{h}_{p,q,k}+\hspace*{-0.5mm}\sigma ^2_{\mathrm{I}_k}}.\\
    \hspace*{-6mm}P^{\mathrm{ER}}_j&\hspace*{-3mm}=\hspace*{-3mm}&\underset{\substack{s,p\in\mathcal{S},\\t,q\in\widehat{\mathcal{T}}}}{\sum }b_{s,t}b_{p,q}\mathbf{g}^H_{s,t,j}\left (  \underset{k\in\mathcal{K}}{\sum }\mathbf{w}_k\mathbf{w}^H_k+\underset{i\in\mathcal{J}}{\sum }\mathbf{v}_i\mathbf{v}^H_i \right )\mathbf{g}_{p,q,j}.\label{HarvestedPower}
\end{eqnarray}
\setcounter{equation}{20}
\begin{eqnarray}
\label{Chat1}
\hspace*{-5mm}&&\hspace*{-6mm}\widehat{\mbox{C1}}\mbox{:}\hspace*{1mm}\frac{1}{\Gamma_k^{\mathrm{req}}}\underset{\substack{s,p\in\mathcal{S},\\t,q\in\widehat{\mathcal{T}}}}{\sum }\mathrm{Tr}\big(\mathbf{h}_{p,q,k}\mathbf{h}^H_{s,t,k} \widehat{\mathbf{W}}_{k,s,p,t,q}\big)-\hspace*{-2mm}\underset{\substack{s,p\in\mathcal{S},\\t,q\in\widehat{\mathcal{T}}}}{\sum }\mathrm{Tr}\big (\mathbf{h}_{p,q,k}\mathbf{h}^H_{s,t,k} \big (\hspace*{-2mm}\underset{r\in\mathcal{K}\setminus \left \{ k \right \} }{\sum }\hspace*{-2mm}\widehat{\mathbf{W}}_{r,s,p,t,q}+\underset{j\in\mathcal{J}}{\sum } \widehat{\mathbf{V}}_{j,s,p,t,q}\big )\big )\geq \sigma ^2_{\mathrm{I}_k},\\[2mm]\notag\\
\hspace*{-5mm}&&\hspace*{-6mm}\widehat{\mbox{C2}}\mbox{:}\hspace*{1mm}C_{\mathrm{req}_j}\geq \mathrm{exp}\big(-\varrho _j\underset{\substack{s,p\in\mathcal{S},\\t,q\in\widehat{\mathcal{T}}}}{\sum }\mathrm{Tr}\big ( \mathbf{g}_{p,q,j} \mathbf{g}^H_{s,t,j}\big (  \underset{k\in\mathcal{K}}{\sum }\widehat{\mathbf{W}}_{k,s,p,t,q}+\underset{i\in\mathcal{J}}{\sum }\widehat{\mathbf{V}}_{i,s,p,t,q}\big )\big ) \big ),\forall j, \label{Chat2}
\end{eqnarray}
\hrule
\end{figure*}
The receive signal-to-noise-plus-interference ratio (SINR) of IR $k$, i.e., $\Gamma_k$ is given in \eqref{SINR} shown at the top of the next page.
Furthermore, the received RF power at ER $j$, i.e., $P_j^{\mathrm{ER}}$, is given in \eqref{HarvestedPower} shown at the top of the next page.
\par
\setcounter{equation}{9}
In this paper, we assume that the transmission mode set is given and we focus on the online resource allocation algorithm design for the considered system. In particular, we aim to minimize the total transmit power at the BS while satisfying the QoS requirements of the IRs and the EH requirements of the ERs. In particular, the joint beamforming and transmission mode selection policy, i.e., $\left \{\mathbf{w}_k,\mathbf{v}_j,b_{s,t} \right \}$, is obtained by solving the following optimization problem
\begin{eqnarray}
\label{prob1}
&&\hspace*{0mm}\underset{\mathbf{w}_k,\mathbf{v}_j,b_{s,t}}{\mino} \,\, \,\, \hspace*{2mm}\underset{k\in\mathcal{K}}{\sum }\left \| \mathbf{w}_k \right \|^2+\underset{j\in\mathcal{J}}{\sum }\left \| \mathbf{v}_j \right \|^2\notag\\
\mbox{s.t.}\hspace*{-4mm}
&&\mbox{C1:~}\Gamma_k\geq \Gamma_{\mathrm{req}_k},~\forall k,\hspace*{9.6mm}\mbox{C2:~}\Upsilon_j^{\mathrm{EH}}\geq E_{\mathrm{req}_j},~\forall j,\notag\\
&&\mbox{C3:~}b_{s,t}\in\left \{ 0,1 \right \},~\forall s,~\forall t,\hspace*{2mm}
\mbox{C4:~}\underset{ s\in\mathcal{S}}{\sum}b_{s,t}=1,~\forall t.
\end{eqnarray}
Here, $\Gamma_{\mathrm{req}_k}$ in constraint C1 is the predefined minimum required SINR of IR $k$. Constraint C2 indicates that the minimum harvested power at ER $j$ should be greater than a given threshold $E_{\mathrm{req}_j}$. Constraints C3 and C4 are imposed since only one transmission mode can be selected for each tile.
\par
We note that the optimization problem in \eqref{prob1} is a non-convex problem. In particular, the non-convexity stems from the coupling of the optimization variables, the fractional function in constraint C1, and the binary selection constraint in C3. In general, it is not possible to solve \eqref{prob1} optimally in polynomial time. Hence, in the next section, we propose a computationally efficient SCA-based suboptimal algorithm to handle the optimization problem.

\section{Solution of the Problem}
Let us first define $\mathbf{W}_k=\mathbf{w}_k\mathbf{w}_k^H$, $\forall k$, and $\mathbf{V}_j=\mathbf{v}_j\mathbf{v}_j^H$, $\forall j$. Also, we define new optimization variable $\beta_{ s,p,t,q} =b_{s,t}b_{p,q}$, $\forall s,p\in\mathcal{S}$, $\forall t,q\in\widehat{\mathcal{T}}$, which satisfies the following convex constraints:
\begin{eqnarray}
&&\hspace*{-14mm}\mbox{C5a:~}0\leq \beta_{ s,p,t,q}\leq 1,\hspace*{5mm}
\mbox{C5b:~}\beta_{ s,p,t,q}\leq b_{s,t},\\
&&\hspace*{-14mm}\mbox{C5c:~}\beta_{ s,p,t,q}\leq b_{p,q},\hspace*{8mm}
\mbox{C5d:~}\beta_{ s,p,t,q}\geq b_{s,t}+b_{p,q}-1.
\end{eqnarray}
Then, the SINR of IR $k$ and the received RF power at ER $j$ can be rewritten as follows, respectively,
\begin{eqnarray}
\hspace*{-4mm}\Gamma_k&\hspace*{-3mm}=\hspace*{-3mm}&\frac{\underset{\substack{s,p\in\mathcal{S},\\t,q\in\widehat{\mathcal{T}}}}{\sum }\beta_{ s,p,t,q}\mathbf{h}^H_{s,t,k} \mathbf{W}_k\mathbf{h}_{p,q,k}}{\hspace*{-3mm}\underset{\substack{s,p\in\mathcal{S},\\t,q\in\widehat{\mathcal{T}}}}{\sum }\hspace*{-1mm}\beta_{ s,p,t,q}\mathbf{h}^H_{s,t,k}\hspace*{-1mm}\left (\hspace*{-0.5mm}  \underset{r\in\mathcal{K}\setminus \left \{ k \right \} }{\hspace*{-3mm}\sum }\hspace*{-4mm}\mathbf{W}_r\hspace*{-0.5mm}+\hspace*{-1mm}\underset{j\in\mathcal{J}}{\sum }\mathbf{V}_j\hspace*{-1mm}\right )\hspace*{-0.5mm}\mathbf{h}_{p,q,k}\hspace*{-0.5mm}+\hspace*{-0.5mm}\sigma ^2_{\mathrm{I}_k}},\\
\hspace*{-4mm}P_j^{\mathrm{ER}}&\hspace*{-3mm}=\hspace*{-3mm}&\underset{\substack{s,p\in\mathcal{S},\\t,q\in\widehat{\mathcal{T}}}}{\sum }\beta_{ s,p,t,q}\mathbf{g}^H_{s,t,j}\left (  \underset{k\in\mathcal{K}}{\sum }\mathbf{W}_k+\underset{i\in\mathcal{J}}{\sum }\mathbf{V}_i\right )\mathbf{g}_{p,q,j}.
\end{eqnarray}
One obstacle to solving the problem in \eqref{prob1} is the coupling of the optimization variables. We overcome this difficulty by applying the big-M formulation \cite{griva2009linear}. In particular, we define new optimization variables $\widehat{\mathbf{W}}_{k,s,p,t,q}=\beta_{ s,p,t,q}\mathbf{W}_k$ and $\widehat{\mathbf{V}}_{j,s,p,t,q}=\beta_{ s,p,t,q}\mathbf{V}_j$ and decompose the product terms by imposing the following additional convex constraints:
\begin{eqnarray}
\hspace*{-16mm}&&\mbox{C6a:~}\widehat{\mathbf{W}}_{k,s,p,t,q}\preceq\beta_{ s,p,t,q}P^{\mathrm{max}}\mathbf{I}_{N_{\mathrm{T}}},\\
\hspace*{-16mm}&&\mbox{C6b:~}\widehat{\mathbf{W}}_{k,s,p,t,q}\succeq \mathbf{W}_k-(1-\beta_{ s,p,t,q})P^{\mathrm{max}}\mathbf{I}_{N_{\mathrm{T}}},\\
\hspace*{-16mm}&&\mbox{C6c:~}\widehat{\mathbf{W}}_{k,s,p,t,q}\preceq \mathbf{W}_k,\hspace*{11mm}\mbox{C6d:~}\widehat{\mathbf{W}}_{k,s,p,t,q}\succeq \mathbf{0},\\
\hspace*{-16mm}&&\mbox{C7a:~}\widehat{\mathbf{V}}_{j,s,p,t,q}\preceq\beta_{ s,p,t,q}P^{\mathrm{max}}\mathbf{I}_{N_{\mathrm{T}}},\\
\hspace*{-16mm}&&\mbox{C7b:~}\widehat{\mathbf{V}}_{j,s,p,t,q}\succeq \mathbf{V}_j-(1-\beta_{ s,p,t,q})P^{\mathrm{max}}\mathbf{I}_{N_{\mathrm{T}}},\\
\hspace*{-16mm}&&\mbox{C7c:~}\widehat{\mathbf{V}}_{j,s,p,t,q}\preceq \mathbf{V}_j,\hspace*{14mm}
\mbox{C7d:~}\widehat{\mathbf{V}}_{j,s,p,t,q}\succeq \mathbf{0},
\end{eqnarray}
where $P^{\mathrm{max}}$ denotes the maximum transmit power allowance available at the BS. Then, constraints C1 and C2 can be recast as constraints $\widehat{\mbox{C1}}$ and $\widehat{\mbox{C2}}$ which are given in \eqref{Chat1} and \eqref{Chat2} and shown at the top of this page, respectively.
Constant $C_{\mathrm{req}_j}$ in $\widehat{\mbox{C2}}$ is defined as $C_{\mathrm{req}_j}=(\frac{a_j}{E_{\mathrm{req}_j}(1-\Xi_j)+a_j\Xi_j}-1)\mathrm{exp}(-\varrho _jc_j)$.
Then, the optimization problem in \eqref{prob1} can be equivalently rewritten as follows
\setcounter{equation}{22}
\begin{eqnarray}
\label{prob2}
&&\hspace*{-12mm}\underset{\substack{\mathbf{W}_k,\widehat{\mathbf{W}}_{k,s,p,t,q}\in\mathbb{H}^{N_{\mathrm{T}}},\\\mathbf{V}_j,\widehat{\mathbf{V}}_{j,s,p,t,q}\in\mathbb{H}^{N_{\mathrm{T}}},\\b_{s,t},\beta_{ s,p,t,q}}}{\mino} \,\, \,\, \mathrm{Tr}\left (\underset{k\in\mathcal{K}}{\sum }  \mathbf{W}_k+\underset{j\in\mathcal{J}}{\sum } \mathbf{V}_j\right )\notag\\
&&\hspace*{-12mm}\mbox{s.t.~}\widehat{\mbox{C1}},\widehat{\mbox{C2}},\mbox{C3},\mbox{C4},\mbox{C5a-C5d},\mbox{C6a-C6d},\mbox{C7a-C7d},\notag\\
&&\hspace*{-6.5mm}\mbox{C8:~}\mathrm{Rank}(\mathbf{W}_k)\leq 1,~\forall k,\hspace*{1mm}\mbox{C9:~}\mathrm{Rank}(\mathbf{V}_j)\leq 1,~\forall j,
\end{eqnarray}
where $\mathbf{W}_k,\mathbf{V}_j\in\mathbb{H}^{N_{\mathrm{T}}}$, and constraints C8 and C9 are imposed to guarantee that $\mathbf{W}_k=\mathbf{w}_k\mathbf{w}_k^H$ and $\mathbf{V}_j=\mathbf{v}_j\mathbf{v}_j^H$ hold after optimization. We note that the non-convex constraints C3, C8, and C9 are still obstacles to solving problem \eqref{prob2}. In particular, constraint C3 is a binary constraint which is intrinsically non-convex. For handling this, 
we first rewrite constraint C3 equivalently as follows:
\begin{eqnarray}
\mbox{C3a:~}\hspace*{-1mm}\underset{\substack{s\in\mathcal{S},t\in\widehat{\mathcal{T}}} }{\sum }\hspace*{-2mm}b_{s,t}-b_{s,t}^2\leq 0~\mbox{and}~
\mbox{C3b:~}0\leq b_{s,t}\leq 1,~\forall s,t.
\end{eqnarray}
Yet, we note that constraint $\mbox{C3a}$ involves a difference of convex functions. Hence, we employ the penalty method \cite{ben1997penalty} and recast \eqref{prob2} as follows:
\begin{eqnarray}
\label{prob3}
&&\hspace*{-7mm}\underset{\substack{\mathbf{W}_k,\widehat{\mathbf{W}}_{k,s,p,t,q}\in\mathbb{H}^{N_{\mathrm{T}}},\\\mathbf{V}_j,\widehat{\mathbf{V}}_{j,s,p,t,q}\in\mathbb{H}^{N_{\mathrm{T}}},\\b_{s,t},\beta_{ s,p,t,q}}}{\mino} \,\, \,\, \hspace*{-4mm}\mathrm{Tr}\left (\underset{k\in\mathcal{K}}{\sum }  \mathbf{W}_k\hspace*{-1mm}+\hspace*{-1mm}\underset{j\in\mathcal{J}}{\sum } \mathbf{V}_j\right )\hspace*{-0.5mm}+\chi\underset{\substack{s\in\mathcal{S},\\t\in\widehat{\mathcal{T}}} }{\sum }(b_{s,t}\hspace*{-0.5mm}-b_{s,t}^2)\notag\\
&&\hspace*{4mm}\mbox{s.t.}\hspace*{6mm}\widehat{\mbox{C1}},\widehat{\mbox{C2}},\mbox{C3b},\mbox{C4},\mbox{C5a-C7d},\mbox{C8},\mbox{C9},
\end{eqnarray}
where $\chi\gg0$ is a constant penalty factor which ensures that $b_{s,t}$ is binary. Next, we reveal the equivalence between problem \eqref{prob3} and problem \eqref{prob2} in the following theorem \cite{ben1997penalty}.
\par
\textit{Theorem 1:~} Denote the optimal solution of problem \eqref{prob3} as $(b_{s,t})_i$ with penalty factor $\chi=\chi_i$. When $\chi_i$ is sufficiently large, i.e., $\chi=\chi_i\rightarrow \infty$, every limit point $(b_{s,t})$ of the sequence $\left \{ (b_{s,t})_i \right \}$ is an optimal solution of problem \eqref{prob2}.
\par
\textit{Proof:~}Theorem 1 can be proved by following a similar approach as in \cite [Appendix B]{xu2020resource} and is omitted here due to the page limitation. \qed
\par 
We note that the objective function of \eqref{prob3} is in the canonical form of a difference of convex programming problem, which facilitates the application of SCA. In particular, for a given feasible point $b_{s,t}^{(m)}$ found in iteration $m$ of the SCA procedure, we construct a global underestimator of $b_{s,t}^2$ as follows
\begin{equation}
    b_{s,t}^2 \geq 2 b_{s,t}b_{s,t}^{(m)}-(b_{s,t}^{(m)})^2,~\forall s,t.
\end{equation}
\par
The optimization problem solved in the $(m+1)$-th iteration of the proposed algorithm is given by:
\begin{eqnarray}
\label{prob4}
&&\hspace*{-10mm}\underset{\substack{\mathbf{W}_k,\widehat{\mathbf{W}}_{k,s,p,t,q}\in\mathbb{H}^{N_{\mathrm{T}}},\\\mathbf{V}_j,\widehat{\mathbf{V}}_{j,s,p,t,q}\in\mathbb{H}^{N_{\mathrm{T}}},\\b_{s,t},\beta_{ s,p,t,q}}}{\mino} \,\, \,\, f(\Omega^{(m)})\notag\\
&&\hspace*{0mm}\mbox{s.t.}\hspace*{4mm}
\widehat{\mbox{C1}},\widehat{\mbox{C2}},\mbox{C3b},\mbox{C4},\mbox{C5a-C7d},\mbox{C8},\mbox{C9},
\end{eqnarray}
where set $\Omega^{(m)}$ contains the intermediate solution in the $m$-th iteration of the developed algorithm and $f(\Omega^{(m)})\overset{\Delta }{=}\mathrm{Tr}\left (\underset{k\in\mathcal{K}}{\sum }  \mathbf{W}_k+\underset{j\in\mathcal{J}}{\sum } \mathbf{V}_j\right )+\chi\underset{\substack{s\in\mathcal{S},\\t\in\widehat{\mathcal{T}}} }{\sum }\Big(b_{s,t}\hspace*{-0.5mm}-2 b_{s,t}b_{s,t}^{(m)}+(b_{s,t}^{(m)})^2\Big)$ is the corresponding objective function value. We note that the only non-convex constraints in \eqref{prob4} are the unit-rank constraints C8 and C9. To overcome this, we omit constraints C8 and C9 by applying SDR. As a result, the rank-relaxed version of \eqref{prob4} becomes a standard convex optimization problem which can be solved by convex program solvers such as CVX \cite{grant2008cvx}. We note that the solution of the relaxed problem is the global optimum of \eqref{prob4}. Next, we show the tightness of the relaxation by introducing the following theorem.
\par
\textit{Theorem 2:~}For given $\Gamma_{\mathrm{req}_k}>0$, the optimal solution $\mathbf{W}_k^*$ and $\mathbf{V}_j^*$ always satisfies $\mathrm{Rank}(\mathbf{W}_k^*)=1$, $\forall k$, and $\mathrm{Rank}(\mathbf{V}_j^*)\leq1$, $\forall j$, respectively.
\par
\textit{Proof:~}Problem \eqref{prob4} has a similar structure as \cite [Problem (69)]{xu2019resource} and Theorem 2 can be proved following the same steps as in \cite [Appendix A]{xu2019resource}. Due to the space constraints, we omit the detailed proof of Theorem 2. \qed
\par
\begin{algorithm}[t]
\caption{SCA-based Algorithm}
\begin{algorithmic}[1]
\STATE Set iteration index $m=1$, initial point $\mathbf{W}_k^{(1)}$, $\mathbf{V}_j^{(1)}$, $\widehat{\mathbf{W}}_{k,s,p,t,q}^{(1)}$, $\widehat{\mathbf{V}}_{j,s,p,t,q}^{(1)}$, $b_{s,t}^{(1)}$, $\beta_{ s,p,t,q}^{(1)}$, and error tolerance $0<\varepsilon\ll1$.
\REPEAT
\STATE For given $\mathbf{W}_k^{(m)}$, $\mathbf{V}_j^{(m)}$, $\widehat{\mathbf{W}}_{k,s,p,t,q}^{(m)}$, $\widehat{\mathbf{V}}_{j,s,p,t,q}^{(m)}$, $b_{s,t}^{(m)}$, $\beta_{ s,p,t,q}^{(m)}$ obtain an intermediate solution by solving the relaxed version of problem \eqref{prob4}
\STATE Set $m=m+1$
\UNTIL $\frac{f(\Omega^{(m-1)})-f(\Omega^{(m)})}{f(\Omega^{(m)})}\leq \varepsilon$
\end{algorithmic}
\end{algorithm}
\par
The overall algorithm is summarized in \textbf{Algorithm 1}. We note that in each iteration of \textbf{Algorithm 1}, the objective function in \eqref{prob3} is monotonically decreasing. Moreover, as $\chi\rightarrow\infty $, the proposed algorithm asymptotically converges to a locally optimal solution of \eqref{prob3} in polynomial time. The per iteration computational complexity of \textbf{Algorithm 1} scales with the number of tiles (but not with the number of IRS elements) and the number of transmission modes and is given by $\mathcal{O}\Big((K+J)S^2(T+1)^2N_{\mathrm{T}}^{3}+\big((K+J)S^2(T+1)^2\big)^2N_{\mathrm{T}}^{2}+\big((K+J)S^2(T+1)^2\big)^3\Big)$, where $\mathcal{O}\left ( \cdot  \right )$ is the big-O notation \cite[Theorem 3.12]{polik2010interior}. 
\section{Simulation Results}
\begin{table}[t]\vspace*{0mm}\caption{System parameters.}\vspace*{0mm}\label{tab:parameters}\footnotesize
\newcommand{\tabincell}[2]{\begin{tabular}{@{}#1@{}}#2\end{tabular}}
\centering
\begin{tabular}{|l|l|l|}\hline
    \hspace*{-1mm}$f_c$ & Carrier center frequency & $2.5$ GHz \\
\hline
    \hspace*{-1mm}$\alpha$ & Path loss exponent & $2$ \\
\hline
    \hspace*{-1mm}$S$ & Number of transmission modes & $15$ \\
\hline
    \hspace*{-1mm}$N_{\mathrm{T}}$ & Number of antennas at the BS & $8$ \\
\hline
    \hspace*{-1mm}$\Gamma_{\mathrm{req}_k}$ & Min. required SINR at IR $k$ & $10$ dB \\ 
\hline
    \hspace*{-1mm}$E_{\mathrm{req}_j}$ & Min. required energy at ER $j$ & $-10$ dBm \\ 
\hline
    \hspace*{-1mm}$G_i$ & BS antenna gain & $5$ dBi \\
\hline
    \hspace*{-1mm}$a_j$& EH parameter & $20$ mW \cite{boshkovska2015practical}\\
\hline
    \hspace*{-1mm}$c_j$, $\varrho _j$ & EH parameters & $6400$, $0.3$ \cite{boshkovska2015practical}\\
\hline
    \hspace*{-1mm}$L_{\mathrm{D}_k}$, $L_{\mathrm{T}}$, $L_{\mathrm{R}_k}$ & Numbers of scatterers & $4$ \\
\hline
    \hspace*{-1mm}$\sigma^2_{\mathrm{I}_k}$, $\sigma^2_{\mathrm{E}_j}$ & Noise power at receivers & $-90$ dBm \\
\hline
    \hspace*{-1mm}$\varepsilon$ & SCA error tolerance & $10^{-3}$ \\
\hline
    \hspace*{-1mm}$\chi$ & Penalty factor & $10^{3}$\\
\hline
\end{tabular}
\end{table}
In this section, we evaluate the performance of the
proposed scheme via simulations. 
In particular, we assume that there are $K=2$ IRs and $J=2$ ERs in the SWIPT system. We consider a rectangular IRS comprising $24\times 20$ elements in the $x$-$y$ plane with an element spacing of half a wavelength. The IRS is $30$ m away from the BS. To facilitate computationally efficient resource allocation algorithm design, we partition the $480$ IRS elements into $T$ tiles of equal size and jointly design all the elements of each tile offline to generate a set of transmission modes. Following a similar approach as in \cite [Section III-A]{najafi2020physics}, for all tiles, we generate identical reflection codebooks with $121$ transmission modes and wavefront phase codebooks with $3$ transmission modes\footnote{The overall codebook designed in the offline stage is the product of the reflection codebook and the wavefront phase codebook. In particular, the reflection codebook enables the tile to reflect an
incident signal with desired elevation and azimuth angles while the wavefront phase codebook facilitates the constructive or destructive combination of the signals that arrive from different tiles at the receivers.}. Then, we adjust the size of the reflection codebook by selecting the $S_r$ transmission modes which yield the strongest channels. In particular, we calculate the Euclidean norm of $\mathbf{h}_{s,t,k}$, i.e., $\left \| \mathbf{h}_{s,t,k} \right \|_2$, and select the $S_r$ channel vectors with the largest magnitude $\left \| \mathbf{h}_{s,t,k} \right \|_2$. As a result, the size of the transmission mode set for online optimization is given by $S=3S_r$, where $S_r\in\left \{ 10, 15, 30, 100 \right \}$ for our simulations. Moreover, we assume the channel coefficients in $\mathbf{C}_{\mathrm{T}}$, $\mathbf{C}_{\mathrm{R}_k}$, and $\mathbf{C}_{\mathrm{D}_k}$ are impaired by free space path loss and Rayleigh fading. Due to the severe shadowing of the direct link, we assume shadowing attenuations of $-30$ dB and $0$ dB for the direct links and the reflected links, respectively. The AoAs and AoDs at the BS and the IRS are uniformly distributed random variables and are generated as follows: The azimuth angles and polarizations of the incident signal are uniformly distributed in the interval $[0,2\pi]$. The elevation angles of the IRS and the BS are uniformly distributed in the range of $[0,\pi/4]$ while the elevation angles of all users are uniformly distributed in the interval $[0,\pi]$. The adopted parameter values are listed in Table \ref{tab:parameters}.
\par
\begin{figure}[t]\vspace*{0mm}
 \centering
\includegraphics[width=3.4in]{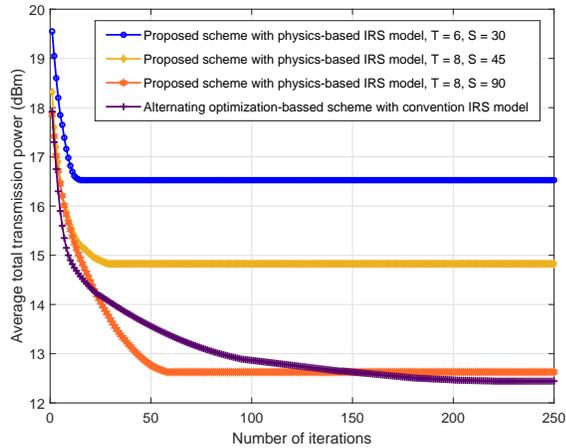}
\vspace*{0mm}
\caption{Convergence for the proposed algorithm of different IRS models for $K=2$, $J=2$, $N_{\mathrm{T}}=8$, $\Gamma_{\mathrm{req}_k}=10$ dB, and $E_{\mathrm{req}_j}=-10$ dBm.}\label{poweriteraation_SWIPT}
\end{figure}
\begin{table}[t]
\caption{Comparison of the average convergence runtime for different schemes.}
\label{tab:runtime}\footnotesize
\newcommand{\tabincell}[2]{\begin{tabular}{@{}#1@{}}#2\end{tabular}}
\centering
\begin{tabular}{|c|c|}\hline
\hspace*{-1mm} Scheme & Avg. runtime (s)\\
\hline
\hspace*{-1mm} Proposed scheme with $T=6$ and $S=30$ & $1845$ \\
\hline
\hspace*{-1mm} Proposed scheme with $T=8$ and $S=45$ & $4220$\\
\hline
\hspace*{-1mm} Proposed scheme with $T=8$ and $S=90$ & $16725$\\
\hline
\hspace*{-1mm} AO-based scheme with conventional IRS model & $24903$\\
\hline
\end{tabular}
\end{table}
In Figure \ref{poweriteraation_SWIPT}, we investigate the convergence of the proposed algorithm for different IRS models. Apart from the problem in \eqref{prob1}, we also apply the AO-based algorithm proposed in \cite{yu2019robust} to a problem similar to \eqref{prob1} but with the conventional IRS model. In particular, for the conventional IRS model, we jointly optimize the phase shifts of all IRS elements as well as the transmit beamformers for minimization of the total transmit power while satisfying the QoS requirements of the IRs and ERs. As can be seen from Figure \ref{poweriteraation_SWIPT}, for a large IRS with $480$ elements, solving the problem for the conventional IRS model requires a considerably larger number of iterations compared to the physics-based IRS model. This is because the physics-based IRS model allows us to decouple the size of the search space of the developed algorithm from the number of IRS elements. Instead, the size of the search space scales with the number of tiles and the size of the transmission mode set. However, by narrowing the search space, the proposed scheme based on the physics-based IRS model also suffers from a performance loss compared to the scheme based on the conventional IRS model. Furthermore, although the proposed schemes with $T=8$ and $S=45$ and with $T=8$ and $S=90$ result in better performances compared to the scheme with $T=6$ and $S=30$, they also require $15$ more and $40$ more iterations to converge, respectively. This reveals that by reconfiguring the tiles and resizing the transmission mode set, the number of iterations required for convergence of \textbf{Algorithm 1} can be controlled to trade complexity and performance. Furthermore, for the parameters adopted in Figure \ref{poweriteraation_SWIPT}, we provide the average convergence runtime\footnote{These simulations were carried out on a computer equipped with an Intel Core i7-3770 processor with a base frequency of $3.40$ GHz.} of the proposed \textbf{Algorithm 1} and the AO-based scheme in Table \ref{tab:runtime} (in seconds). As can be observed from Table \ref{tab:runtime}, the AO-based scheme with the conventional IRS model requires a significantly larger runtime compared to the proposed scheme with the physics-based IRS model indicating a higher complexity and slower convergence. 
\par
\begin{figure}[t]\vspace*{0mm}
 \centering
\includegraphics[width=3.4in]{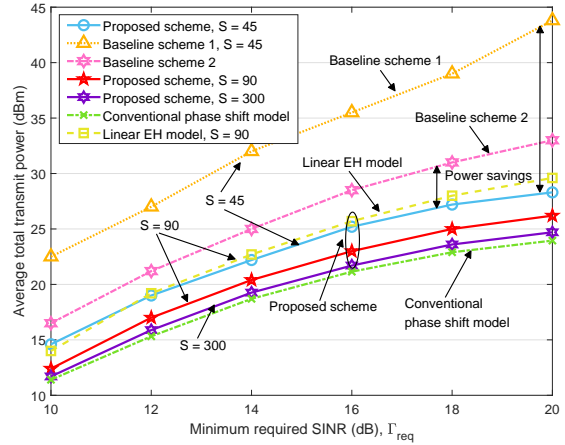}
\vspace*{0mm}
\caption{Average total transmit power (dBm) versus minimum required SINR at IRs for different resource allocation schemes with $K=2$, $J=2$, $T=8$, $N_{\mathrm{T}}=8$, and $E_{\mathrm{req}_j}=-10$ dBm.}\label{powerSINR_SWIPT}
\end{figure}
In Figure \ref{powerSINR_SWIPT}, we investigate the average total transmit power versus the minimum required SINR at the IRs, $\Gamma_{\mathrm{req}}=\Gamma_{\mathrm{req}_k}$, $\forall k$, for different resource allocation schemes. To investigate the effectiveness of the algorithm developed in this paper, we consider two baseline schemes. For baseline scheme 1, we randomly assign a transmission mode to each tile, and the BS employs maximum ratio transmission where each beamformer is aligned to the corresponding channel vector. Then, the transmit power at the BS is minimized by optimizing the power allocated to each user. For baseline scheme 2, we evaluate the system performance when an IRS is not deployed. Then, we optimize the beamforming vectors for minimization of the transmit power. As can be observed from Figure \ref{powerSINR_SWIPT}, the required total transmit powers of the proposed scheme and the two baseline schemes grow with $\Gamma_{\mathrm{req}}$. Moreover, the proposed scheme achieves significant power savings compared to the two baseline schemes. This reveals the effectiveness of the proposed scheme for jointly optimizing the beamforming vectors and transmission mode selection. Furthermore, we also study the transmit power of the proposed scheme for different transmission mode set sizes. In particular, as the size of the transmission mode set increases from $45$ to $90$, the BS consumes less power to meet the QoS requirements of the IRs and ERs. This is due to the fact that for a larger transmission mode set size, the IRS can perform more precise beamforming which results in power savings. However, as we further increase the size of the transmission mode set to $300$, the additional power savings of the proposed scheme are limited. In fact, the proposed scheme with $S=300$ transmission modes closely approaches the performance of the AO-based scheme employing the conventional phase shift model where all the IRS elements are jointly optimized. We note that, for an even larger IRS (e.g. more than $1000$ phase shift elements), the AO-based scheme with the conventional IRS model would be prohibitively complex, while for the proposed scheme the values of $T$ and $S$ can still be properly set for efficient online optimization. On the other hand, we also evaluate the performance of the proposed scheme if the conventional linear EH model is adopted. In particular, we solve a problem similar to \eqref{prob1} but with the linear EH model. Then, we employ the obtained solution in the actual system with non-linear EH and check if the QoS requirements of the ERs are satisfied. In fact, due to the non-linear nature of the EH circuits, the proposed scheme with the linear EH model may cause saturation in the EH circuit of some of the ERs and underutilization of other ERs. If the QoS requirements of the ERs are not satisfied, we increase the power allocated to the ERs until constraint C2 in \eqref{prob1} is fulfilled. As can be observed from Figure \ref{powerSINR_SWIPT}, to satisfy the QoS requirement of the ERs, the scheme based on the linear EH model consumes more power compared to the proposed scheme based on the non-linear EH model. 
\vspace*{2mm}
\section{Conclusion}
\vspace*{2mm}
In this paper, we investigated the resource allocation algorithm design for large IRS-assisted SWIPT systems. Compared to existing works assuming an overly simplified system model, we adopted a non-linear EH model and a physics-based IRS model which better reflect the properties of practical IRS-assisted SWIPT systems and lead to scalable optimization for large IRSs. The required transmission mode set was generated in an offline design stage, and we focused on the joint online optimization of the transmit beamformers and the transmission mode selection for minimization of the total BS transmit power while satisfying the QoS requirements of the IRs and ERs. To tackle the formulated non-convex optimization problem, we developed an SCA-based computationally efficient algorithm which asymptotically converges to a locally optimal solution. Simulation results showed that the proposed scheme achieved considerable power savings compared to two baseline schemes. Moreover, our results revealed that the adopted physics-based channel model and the proposed scalable optimization framework for large IRSs did not only facilitate efficient resource allocation for IRS-assisted SWIPT systems but also enabled a flexible tradeoff between computational complexity and performance. 
\bibliographystyle{IEEEtran}
\bibliography{Reference_List}
\end{document}